\documentclass{PoS}

\title{Gluon Sivers Function and Single Spin Asymmetry in {\bf $e+p^\uparrow \rightarrow e+J/\psi + X$}}

\ShortTitle{Single spin asymmetry in $e+p^\uparrow \rightarrow e+J/\psi + X$}
\author{{Rohini M. Godbole}%
         {}\\
       Centre for High Energy Physics, Indian Institute of Science, Bangalore, India.\\
       E-mail: \email{rohini@cts.iisc.ernet.in}}
\author{{Anuradha Misra}%
        {}\\
       Department of Physics, University of Mumbai, Mumbai, India.\\
       E-mail: \email{misra@physics.mu.ac.in}}
\author{{Asmita Mukherjee}%
        {}\\
       Department of Physics, Indian Institute of Technology Bombay, Mumbai, India.\\
       E-mail: \email{asmita@phy.iitb.ac.in}}
\author{\speaker{Vaibhav S. Rawoot}%
        {}\\
       Department of Physics, University of Mumbai, Mumbai, India.\\
       E-mail: \email{vaibhavrawoot@gmail.com}}

\abstract{We propose measurement of transverse  single spin asymmetry (SSA) in 
 charmonium production as a probe of gluon Sivers function. We estimate 
 SSA in low virtuality electroproduction of $J/\psi $ using color evaporation model 
 of charmonium production and existing models of the gluon Sivers function and 
 find sizable asymmetry at  JLab, HERMES, COMPASS and eRHIC energies.}
 
\FullConference{Sixth International Conference on Quarks and Nuclear Physics\\
                April 16-20, 2012\\
                Ecole Polytechnique, Palaiseau,  Paris}
\newcommand{\be}{\begin{equation}}
\newcommand{\ee}{\end{equation}}
\newcommand{\bea}{\begin{eqnarray}}
\newcommand{\eea}{\end{eqnarray}}

\newcommand{\pup}{p^\uparrow}

\newcommand{\kt}{k_\perp}

\def\lsim{\mathrel{\rlap{\lower4pt\hbox{\hskip1pt$\sim$}}\raise1pt\hbox{$<$}}}
\def\gsim{\mathrel{\rlap{\lower4pt\hbox{\hskip1pt$\sim$}}\raise1pt\hbox{$>$}}}

\begin{document}
%%%%%%%%%%%%%%%%%%%%%%%%%%%%%%%%%%%%%%%%%%%%%%%%%%%%%%%%%%%%%%%%%%%%%%%%%%%%%%%%%%%%%%%%%%%%%%%%%%%
\section{INTRODUCTION}
%%%%%%%%%%%%%%%%%%%%%%%%%%%%%%%%%%%%%%%%%%%%%%%%%%%%%%%%%%%%%%%%%%%%%%%%%%%%%%%%%%%%%%%%%%%%%%%%%%%

Transverse Single Spin Asymmetry (SSA) arises in the scattering of a transversely polarized proton 
off an unpolarised hadron or nucleon if the scattering cross section depends on the 
direction of polarization. The  Single Spin Asymmetry for inclusive process $A^\uparrow + B \rightarrow C+X$ 
is defined as
\be 
A_N = \frac{d\sigma ^\uparrow \, - \, d\sigma ^\downarrow}
{d\sigma ^\uparrow \, + \, d\sigma ^\downarrow} \label{an}
\ee 
where $ d\sigma^{\uparrow(\downarrow)} $ denotes the cross section
for scattering of a transversely  polarized hadron A off an unpolarized hadron B,
with A upwards (downwards) transversely polarized w.r.t. the production plane.
SSA's significantly different from zero 
have been observed over  last 35 years starting with  pion production in scattering of polarized 
protons  off unpolarised proton target~\cite{Klem:1976ui}.
Large SSA's have been measured in  pion production at Fermilab \cite{AdamsBravar1991} as well as 
at BNL-RHIC in $p p^\uparrow $ collisions~\cite{KruegerAllogower1999}. 
SSA's have also been observed by the HERMES \cite{Hermes} and COMPASS \cite{Compass} collaborations, 
in polarized semi-inclusive deep inelastic scattering (SIDIS). 
The magnitude of the observed asymmetries has been found to be larger than
what is predicted by perturbative quantum chromodynamics (pQCD) \cite{alesio-review}. 

Theoretically there are two major approaches to explain the SSA's. 
One is the twist three approach and other is the transverse momentum dependent (TMD) approach which we have used in the present  work.
The TMD approach is based on a pQCD factorization scheme in which spin and intrinsic transverse 
momentum effects are included in parton distribution functions (pdf's) and fragmentation functions (ff's).
One of the difficulties in getting information about the spin and transverse momentum dependent pdf's and 
ff's is that very often two or more of these functions contribute to the same physical observable 
making it difficult to estimate each single one separately. 
% According to this approach SSA's arises due to spin orbit inteaction and therefor can be used as 
% SSA's may involve one or more of these functions. Very often two or more of these functions
% contribute to the same physical process. 
% In the process that we are considering at leading order (LO) there is contribution 
% from only one partonic subprocess $\gamma g \rightarrow c\bar c$ therefore it can be used as a clean probe of gluon Sivers function. 

% Here we are interested in transverse momentum dependent Sivers function  which 
% give the probability of finding an unpolarised partons inside a transversely polarized hadron.
The study of spin asymmetries requires extension of TMD factorization scheme to polarized case.
Sivers in early 90's proposed that there exists a correlation between the azimuthal distribution of an 
unpolarized parton and spin of its parent hadron~\cite{Sivers1990}. 
Number density of partons inside proton with transverse polarization {S}, 
three momentum $\bf p$ and intrinsic transverse momentum $\kt$ of partons
is expressed in terms of Sivers function $\Delta^N f_{a/\pup}(x, k_{\perp})$
\be
\hat f_{a/\pup}(x, {\bf k}_{\perp}, {\bf S}) = \hat f_{a/p}(x, k_{\perp})
+ \frac{1}{2} \, \Delta^N f_{a/\pup}(x, k_{\perp}) \> 
{{\bf S} \cdot (\hat{\bf p} \times \hat{\bf k}_{\perp})}
\ee
${\bf S} \cdot (\hat{\bf p} \times \hat{\bf k}_{\perp})$ gives the correlation 
between the spin of the proton and intrinsic transverse momentum of the unpolarised quarks and gluons.
There have been studies on the quark Sivers function in SIDIS and the gluon Sivers function 
in the process $p^\uparrow p \rightarrow D X$~\cite{kp09, Anselmino2004}. In this work,  
we propose charmonium electroduction as another probe of the gluon Sivers function.

\section{FORMALISM FOR ASYMMETRY IN $J/\psi$ PRODUCTION}

We have estimated SSA in photoproduction  of charmonium in the process 
$e+p^\uparrow \rightarrow e+J/\psi + X$. At leading order (LO), there is contribution only from a 
subprocess $\gamma g \rightarrow c\bar c$. In addition, since we are using color evaporation model (CEM) 
for charmonium production, only one pdf is involved. Thus the process under consideration can be used as 
a clean  probe of the gluon Sivers function. Also since the charmonium production mechanism 
can have implications for this SSA, its study can help throw some light on  the production 
mechanism of charmonium as well.    

% Charmonium production include the two distinct steps, production of $c\bar c$ 
% pair at short distance and its subsequent evolution in to physical charmonium state. The subsequent evolution of $c \bar c$
Charmonium production process can be understood in terms of two distinct steps-
production of a $c\bar c$ pair (a short distance process) and a subsequent binding 
of this pair in to charmonium (a long distance process). Various methods to describe this non-perturbative 
evolution of the $c\bar c$ pair into charmonium lead to different models of charmonium production.
As a first step in our investigations of SSA in charmonium production, we have used the Color Evaporation model (CEM)
of charmonium production. According to CEM,  the cross section for charmonium production is proportional to the
rate of production of $c\bar{c}$ pair integrated over the mass range $2m_c$ to $2m_D$ \cite{cem0, fri, ce2}
\be
\sigma=\frac{1}{9}\int_{2m_c}^{2m_D} dM_{c\bar{c}} \frac{d\sigma_{c\bar{c}}}{dM_{c\bar{c}}}
\label{sigmacem}
\ee  
where $m_c$ is the charm quark mass and $2m_D$ is the $D\bar{D}$ threshold.

The cross section for the low virtuality electroduction within CEM is
\be
{\sigma^{ep\rightarrow e+J/\psi+X}=
\int_{4m_c^2}^{4m_D^2} dM^2_{c\bar{c}} \int dy\> dx\> f_{\gamma/e} (y)\> f_{g/p}(x) 
\>\frac{d\hat{\sigma}^{\gamma g\rightarrow c\bar{c}}}{dM_{c\bar{c}}^2}
\label{xsec-gammap}}
\ee 
where $f_{\gamma/e} (y)$ is the distribution function of the photon in the electron which, 
in the Weizsaker  William approximation~\cite{ww}, is given by 
\bea
f_{\gamma/e}(y,E)=\frac{\alpha}{\pi} \{\frac{1+(1-y)^2}{y}\left(ln\frac{E}{m}-\frac{1}{2}\right)
+\frac{y}{2}\left[ln\left(\frac{2}{y}-2\right)+1\right] \nonumber \\
+\frac{(2-y)^2}{2y}ln\left(\frac{2-2y}{2-y}\right) \}.
\label{ww-function}
\eea

To calculate  SSA in scattering of electrons off a polarized proton target,
we assume generalization of this CEM expression for low virtuality electroproduction of $J/\psi$ by 
taking into account the transverse momentum dependence of the Weizsacker-Williams (WW) 
function and the gluon distribution function: 
\bea
\sigma^{e+p^\uparrow\rightarrow e+J/\psi + X}=
\int_{4m_c^2}^{4m_D^2} dM_{c\bar c}^2\> dx_\gamma\> dx_g\> [d^2{\bf k}_{\perp\gamma}d^2{\bf k}_{\perp g}]\>
f_{g/p^{\uparrow}}(x_{g},{\bf k}_{\perp g}) \nonumber \\
\times f_{\gamma/e}(x_{\gamma},{\bf k}_{\perp\gamma})\>
\frac{d\hat{\sigma}^{\gamma g\rightarrow c\bar{c}}}{dM_{c\bar c}^2}. 
\label{dxec-ep}
\eea

We assume $\kt$ dependence of pdf's and WW function to be factorized in Gaussian form~\cite{kp09} 
%  (Anselmino etal. Eur. Phys. J. A 39, 89 (2009))
\be
f(x,k_{\bot})=f(x)\frac{1}{\pi\langle k^{2}_{\bot}\rangle} 
e^{-k^{2}_{\bot}/\langle{k^{2}_{\bot}\rangle}} 
\label{gauss}
\ee 
with $\langle k^{2}_{\bot}\rangle=0.25~GeV^2.$

The expression for the numerator of the asymmetry is
\bea
\frac{d^{4}\sigma^\uparrow}{dy\>d^2{\bf q}_T}-\frac{d^4\sigma^\downarrow}{dy\>d^2{\bf q}_T}=
\frac{1}{2}\int_{4m^2_c}^{4m^2_D}dM^{2}\>\int [dx_{\gamma}\>dx_{g}\>d^2{\bf k}_{\perp\gamma}\> d^2{\bf k}_{\perp g}]\>
{\Delta^{N}f_{g/p^{\uparrow}}(x_{g},{\bf k}_{\perp g})} \nonumber \\ 
\times f_{\gamma/e}(x_{\gamma},{\bf k}_{\perp\gamma})\> 
\delta^{4}(p_{g}+p_{\gamma}-q)\>
\hat\sigma_{0}^{\gamma g\rightarrow c\bar{c}}(M^2)
\label{nssa}
\eea 
where $q=p_c+p_{\bar c}$, $\Delta^{N}f_{g/p^{\uparrow}}(x_{g},{\bf k}_{\perp g})$
is the gluon Sivers function and $M^2$ is invariant mass of the $c\bar c$ pair.   

The partonic cross section is~\cite{gr78}
\be
\hat{\sigma_{0}}^{\gamma g\rightarrow c\bar{c}}(M^2)=
\frac{1}{2}e_{c}^2\frac{4\pi\alpha\alpha_s}{M^2}
[(1+\gamma-\frac{1}{2}\gamma^2)\ln{\frac{1+\sqrt{1-\gamma}}{1-\sqrt{1-\gamma}}}
-(1+\gamma)\sqrt{1-\gamma}]
\ee
where $\gamma=4\>m_c^2/M^2$.

Sivers asymmetry integrated over the azimuthal angle of $J/\psi$ with a weight factor  
$\sin({\phi}_{q_T}-\phi_S)$ is defined as
% \be
% A_N^{\sin({\phi}_{q_T}-\phi_S)} =\frac{\int d\phi_{q_T}
% [d\sigma ^\uparrow \, - \, d\sigma ^\downarrow]\sin({\phi}_{q_T}-\phi_S)}
% {\int d{\phi}_{q_T}[d{\sigma}^{\uparrow} \, + \, d{\sigma}^{\downarrow}]}
% \label{weight-ssa}
% \ee
\be
A_N=\frac{\int d\phi_{q_T}[\int_{4m^2_c}^{4m^2_D}[dM^{2}]\int[d^2{\bf k}_{\perp g}]
\Delta^{N}f_{g/\pup}(x_{g},{\bf k}_{\perp g})
f_{\gamma/e}(x_{\gamma},{\bf q_T}-{\bf k}_{\perp g})
\hat\sigma_{0}]sin(\phi_{q_T}-\phi_S)}
{2\int d\phi_{q_T}[\int_{4m^2_c}^{4m^2_D}[dM^{2}]\int[d^{2}{\bf k}_{\perp g}]
f_{g/P}(x_g,{\bf k}_{\perp g})
f_{\gamma/e}(x_{\gamma},{\bf q}_T-{\bf k}_{\perp g})
{\hat\sigma}_0]}
\label{an2}
\ee 
where $\phi_{q_T}$ and $\phi_S$ are azimuthal angles of $J/\psi$ and proton spin respectively
and $x_{g,\gamma} = \frac{M}{\sqrt s} \, e^{\pm y}$.

\section{MODELS FOR SIVERS FUNCTION}
We have used the following parameterization for the gluon Sivers function~\cite{kp09}
\be
 \Delta^Nf_{g/\pup}(x,\kt) = 2\,{{\mathcal N}_g(x)}\,h(\kt)\,f_{g/p}(x)
\frac{e^{-k^{2}_{\bot}/\langle{k^{2}_{\bot}\rangle}}}{\pi\langle k^{2}_{\bot}\rangle} 
\cos{\phi_{k_\perp}} .
\label{dnf}
\ee

There is no information available about the gluon Sivers function from experimental data.
The valance and sea quark Sivers distribution functions used are the ones extracted from the 
HERMES and COMPASS experimental data in SIDIS processes~\cite{Anselmino-PRD72}.

The $x$ dependent normalization for u and d quarks is given by,
\be 
{\mathcal N}_f(x) = N_f x^{a_f} (1-x)^{b_f} \frac{(a_f + b_f)^{(a_f +
b_f)}}{{a_f}^{a_f} {b_f}^{b_f}} 
\label{siversx} 
\ee
where $a_f, b_f, N_f$ and $M_1$ are best fit parameters obtained by fitting SIDIS, HERMES and COMPASS data~\cite{kp09}. 

For ${\mathcal N}_g(x)$, we have used two choices~\cite{Boer-PRD69(2004)094025}

\begin{itemize}
\item [(a)] ${\mathcal N}_g(x)=\left( {\mathcal N}_u(x)+{\mathcal N}_d(x) \right)/2 \;$.
\item [(b)] ${\mathcal N}_g(x)={\mathcal N}_d(x)$.
\end{itemize}

For $h(\kt)$, we have used following two choices proposed by Anselmino etal~\cite{kp09, Anselmino2004}:

\begin{itemize}
 \item Model I
\be
{h({\kt}) = \sqrt{2e}\,\frac{{\kt}}{M_{1}}\,e^{-{{\kt}^2}/{M_{1}^2}}},\nonumber
\label{siverskt}
\ee
\item Model II
% \item [Model(2)]\quad\quad\quad
\be
{h({\kt})=\frac{2{\kt} M_0}{{\kt}^2+M_0^{2}}}, \nonumber
\ee
\end{itemize}

where $M_0=\sqrt{\langle{k_\perp^2}\rangle}$ and $M_1$ are best fit parameters.

\section{NUMERICAL ESTIMATES}
We have used the following best fit parameters from the recent HERMES and COMPASS data~\cite{2011-para}
\bea 
N_u = 0.40, \ a_u=0.35, \ b_u =2.6 \; , \nonumber \\
N_d = -0.97, \ a_d = 0.44, \ b_d=0.90 \;, \nonumber \\
M_1^2=0.19~GeV^2.
\eea
% PUT A HEADING - NUMERICAL ESTIMATES - HERE. 

In figure 1 we have shown the comparison of y and $q_T$ distribution of estimated SSA at JLab, 
HERMES, COMPASS and eRHIC for model I and parameterization (a) of the gluon Sivers function. The estimates are
obtained using GRV98LO for gluon distribution function and Weizsaker-Williams function for photon distribution.
The results for model II and parametrization (b) are given in reference\cite{Godbole:2012bx}.  
The hard scale involved in the calculation for all experiments is between $4 m^2_c$ and $4 m^2_D$  as 
we are using color evaporation model. Hence the 
scale evolution of TMD's is not expected to affect much our estimates for the experiments at higher energies.   

According to our estimates sizable asymmetry is expected  at various experiments covering different kinematical regions. Hence it is 
worthwhile to look at SSA's in charmonium production both from the point of view of comparing
different models of charmonium production as well as comparing the different models of gluon 
Sivers function.    

\begin{figure}its
% \hspace*{-2cm}
\includegraphics[width=0.44\linewidth,angle=0]{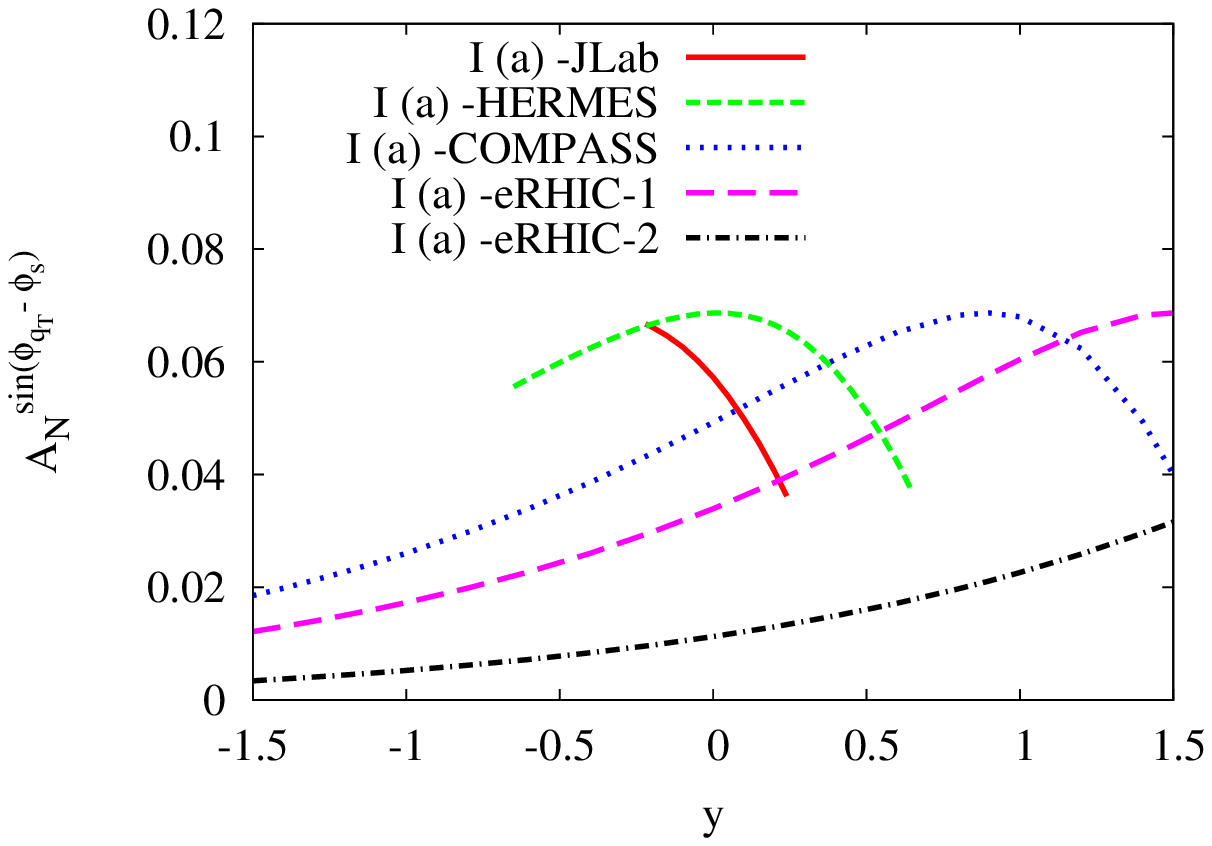}\hspace*{0.2cm} 
\includegraphics[width=0.44\linewidth,angle=0]{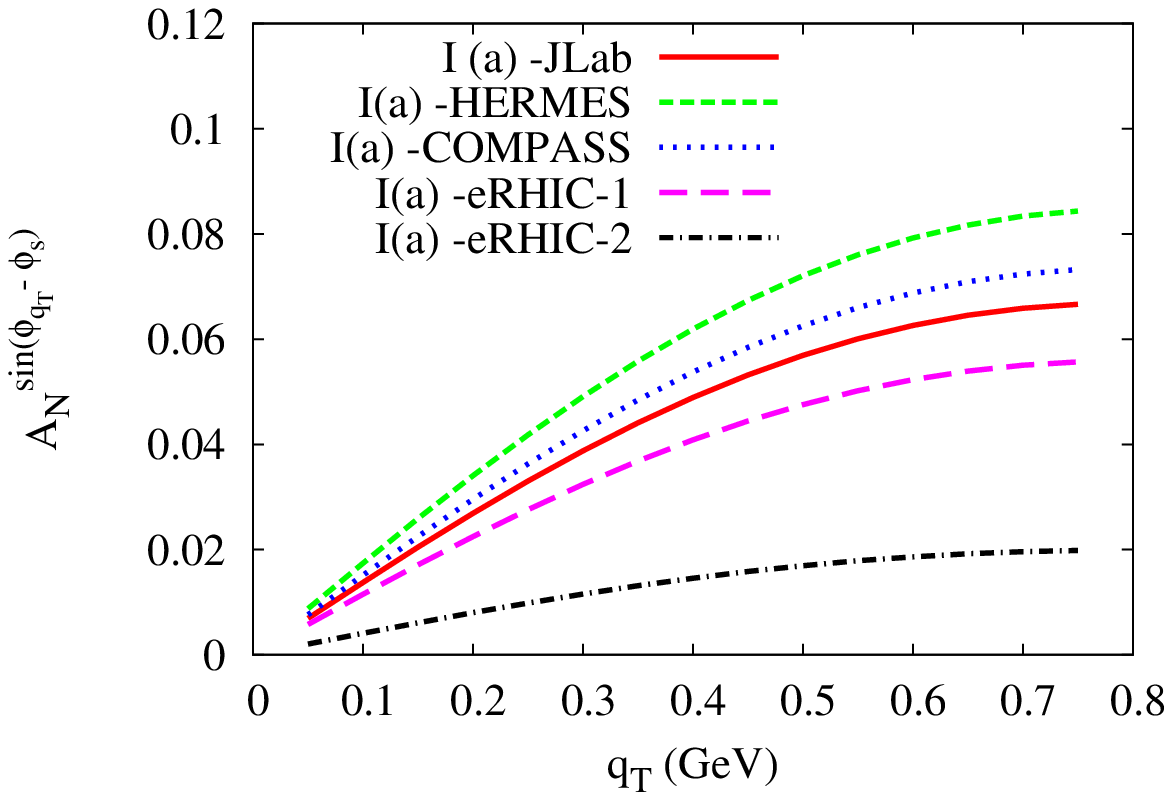}
 \caption{(Color online) The single spin asymmetry $A_N^{\sin({\phi}_{q_T}-\phi_S)}$
for the $e + p^\uparrow \rightarrow e+J/\psi+X$ as a function of y (left panel)  and $q_T$ (right panel). 
The plots are for model I with parameterization (a) compared for JLab ($\sqrt s = 4.7$~GeV) [solid red line], 
HERMES ($\sqrt s = 7.2$~GeV) [dashed green line], COMPASS ($\sqrt s = 17.33$~GeV) [dotted blue line], 
eRHIC-1 ($\sqrt{s}=31.6$~GeV) [long dashed pink line] and eRHIC-2 ($\sqrt{s}=158.1$~GeV) [dot-dashed black line].}  
\end{figure}

%%%%%%%%%%%%%%%%%%%%%%%%%%%%%%%%%%%%%%%%%%%%%%%%%%%%%%%%%%%%%%%%%%%%%%%%%%%%%%%%%%%%%%
\section{ACKNOWLEDGEMENT}
%%%%%%%%%%%%%%%%%%%%%%%%%%%%%%%%%%%%%%%%%%%%%%%%%%%%%%%%%%%%%%%%%%%%%%%%%%%%%%%%%%%%%%
V.S.R. would like to thank organizers of Sixth International Conference on Quarks and Nuclear Physics 2012 (QNP 2012)
for partial financial support and Department of Science and Technology, India for travel support under the 
Grant No. SR/ITS/0041/2012-2013. 
R.M.G. wishes to acknowledge support from the Department of Science and
Technology, India under Grant No. SR/S2/JCB-64/2007. A. Misra and V.S.R. would like to 
thank Department of Science and Technology, India for financial support under the Grant No. 
SR/S2/HEP-17/2006 and the Department of Atomic Energy-BRNS, India under the Grant No. 2010/37P/47/BRNS.

\end{document}